\documentclass[aps,pra,twocolumn,notitlepage,superscriptaddress]{revtex4-1}
\usepackage{bm} \usepackage{graphicx} \usepackage{amsmath}
\usepackage{amsthm,stmaryrd}
\usepackage{amssymb} \usepackage{physics} \usepackage{txfonts}
\usepackage{color}
\usepackage{comment}
\usepackage{hyperref}

\newcommand{\mb}{\boldsymbol}
\newcommand{\eqdef}{:=}

\begin{document}

\title{ Noise-Resilient Variational Hybrid Quantum-Classical Optimization }

\author{Laura Gentini}
\address{Dipartimento di Fisica e Astronomia, Universit\`a di Firenze, I-50019, Sesto Fiorentino (FI), Italy}
\address{INFN, Sezione di Firenze, I-50019, Sesto Fiorentino (FI), Italy}
\author{Alessandro Cuccoli}
\address{Dipartimento di Fisica e Astronomia, Universit\`a di Firenze, I-50019, Sesto Fiorentino (FI), Italy}
\address{INFN, Sezione di Firenze, I-50019, Sesto Fiorentino (FI), Italy}
\author{Stefano Pirandola}
\address{Computer Science and York Centre for Quantum Technologies, University of York,
York YO10 5GH, UK}
\author{Paola Verrucchi}
\address{ISC-CNR, UOS Dipartimento di Fisica, Universit\`a di Firenze, I-50019, Sesto Fiorentino (FI), Italy}
\address{Dipartimento di Fisica e Astronomia, Universit\`a di Firenze, I-50019, Sesto Fiorentino (FI), Italy}
\address{INFN, Sezione di Firenze, I-50019, Sesto Fiorentino (FI), Italy}
\author{Leonardo Banchi}
\address{Dipartimento di Fisica e Astronomia, Universit\`a di Firenze, I-50019, Sesto Fiorentino (FI), Italy}
\address{INFN, Sezione di Firenze, I-50019, Sesto Fiorentino (FI), Italy}
\date{\today}

\begin{abstract}
	Variational hybrid quantum-classical optimization represents one of the most 
	promising avenue to show the advantage of nowadays noisy intermediate-scale quantum computers 
	in solving hard problems, such as finding the minimum-energy state of a Hamiltonian 
	or solving some machine-learning tasks. 
	In these devices noise is unavoidable and impossible to error-correct, yet its
	role in the optimization process is not well understood, especially 
	from the theoretical viewpoint.  Here we consider a minimization problem 
	with respect to a variational state, iteratively obtained via a parametric
	quantum circuit, taking into account both the role of noise and the
	stochastic nature of quantum measurement outcomes. 
	We show that the accuracy of the result obtained for a fixed number of iterations is bounded 
	by a quantity related to the Quantum Fisher Information of the variational state. 
	Using this bound, we study the convergence property of the quantum approximate optimization 
	algorithm under realistic noise models, showing the robustness of the algorithm 
	against different noise strengths. 
\end{abstract}

\maketitle

\section{Introduction} Quantum computers are 
physical devices that are expected to perform calculations essentially 
impossible for our best classical supercomputers 
\cite{arute2019quantum}, although the {\it quantum advantage} has been 
proven only for a specifically designed problem whose practical 
application is unclear.
As for the hardware, the devices that are currently being
built are noisy intermediate-scale quantum (NISQ) ones
\cite{preskill2018quantum}, for which many of the most promising uses 
consist in solving optimization problems via hybrid quantum-classical 
algorithms that include parametric quantum circuits
\cite{peruzzo2014variational,farhi2014quantum,mitarai2018quantum,benedetti2019parameterized,larose2019variational,khatri2019quantum,yuan2019theory,schuld2018circuit}. 
In these algorithms the manipulation of the quantum register 
is done via gates that depend on some parameters:
these are iteratively updated via a feedback 
strategy, where the measurement outcomes of the device are classically 
processed to propose better parameters, in the spirit of a variational 
approach. In what follows, we will refer to the above procedure as a
``hybrid variational optimization''.

There are various aspects that make a real quantum device different from 
an ideal one, amongst which the noise due to any external environment 
and the stochasticity of outcomes, due to the probabilistic nature of 
the quantum measurement process. Different authors, see for instance 
Refs.~\cite{wang2018quantum,mbeng2019quantum,schuld2018circuit,sharma2019noise}, 
studied the effect of noise (e.g.\;noisy gates, dephasing etc.) in 
protocols designed for the noiseless case, and found that noise is 
usually detrimental.
At the same time, the role of stochasticity of outcomes has been described 
using the stochastic gradient descent framework 
\cite{harrow2019low,sweke2019stochastic}. However, how to tame 
the combined effect of noise and stochasticity in hybrid variational 
optimization is still far from being understood.

In this work, we analytically study the convergence properties 
of hybrid variational optimizations in terms of the number of times, 
hereafter dubbed iterations, that the NISQ device must be queried to 
find the optimal parameters with a desired accuracy.
We focus on the effects of noisy gates and stochasticity of the 
measurements outcomes, without considering the further problem of 
choosing the measured observable that is best-suited to extract 
information from the noisy process.
We show that the convergence speed is typically unaffected by the presence of a 
small amount of noise, while the accuracy of the solution typically 
deteriorates as the noise strength increases. 
Moreover, we demonstrate that the attainable accuracy for a fixed 
number of iterations is bounded by a quantity that plays a very relevant 
role in quantum estimation theory, namely the Quantum Fisher 
Information
\cite{braunstein1994statistical,paris2009quantum,giovannetti2011advances} 
Our theoretical prediction is corroborated by the results of numerical 
experiments. 

The paper is structured as follows:
In Sec.~\ref{s:varopt} we introduce the variational hybrid optimization 
procedure and describe an algorithm that implements it. In 
Sec.~\ref{s:bounds} we study the role of noisy operations and 
demonstrate some general results about their effects.
Results of our numerical experiments are presented and discussed
in Sec.~\ref{s:numerics}, while conclusions are drawn in 
Sec.~\ref{s:conclusion}.

\section{Variational Hybrid Optimization} \label{s:varopt}
Consider the expectation value 
\begin{equation}
	C(\mb\theta) \eqdef \bra{\psi(\mb\theta)}\hat H\ket{\psi(\mb\theta)}~,
	\label{cost}
\end{equation}
where $\ket{\psi(\mb\theta)}$ is a quantum state of $N$ qubits
that depends on $P$ classical parameters 
$\mb\theta=(\theta_1,\dots,\theta_P)\in 
\mathbb{R}^P$, and $\hat H$ is a hermitian operator. Many 
optimization problems have their solution in the minimization of 
$C(\mb\theta)$ w.r.t. the parameters $\mb\theta$. When this minimization 
is obtained via a variational procedure,
the state $\ket{\psi(\mb\theta)}$ is dubbed ``variational quantum state''.
As for the operator $\hat H$, its explicit form depends on the specific 
problem under analysis.
In the ``variational quantum eigensolver''
\cite{peruzzo2014variational}, for instance,
$\hat H$ is the Hamiltonian of a quantum many-body system 
and the task is that of finding a good variational approximation of the 
ground state.
In the ``quantum approximate optimization algorithm''
(QAOA) \cite{farhi2014quantum}
the task is that of solving some combinatorial problem, with 
$\hat H$ an Ising-like Hamiltonian whose ground state embodies the 
solution of the problem \cite{lucas2014ising}.
Further, it is possible to express in this language 
some machine learning applications, such as quantum classifiers 
\cite{schuld2018circuit,mcclean2018barren}.
In all of these cases, the function to be minimized, $C(\mb\theta)$, is 
dubbed ``cost'' function and only needs being bound from below, with a 
range of attainable values that depends on the actual physical status of 
the observable represented by the operator $\hat H$.
There also exist optimization instances where $C(\mb\theta)$ can 
have a slightly different meaning. In quantum control 
\cite{khaneja2005optimal} 
and simulation \cite{georgescu2014quantum}, for example, the goal is 
often that of obtaining a quantum state, or quantum operation, that is 
as similar as possible to a given target one;
in this case, one can choose
$\hat H=\hat U\ket{\varphi}\!\bra{\varphi}\hat U^\dagger$,
with $\hat U$ a target unitary and $\ket{\varphi}$ 
a reference state: the function \eqref{cost} is then the square of the 
fidelity of state-preparation, ranging from $0$ to $1$ by definition, 
and $1-C(\mb\theta)$ the function to be minimized.
In this work we will not explicitely refer to this case, as done 
elsewhere \cite{beer2019efficient}, but rather 
focus on problems where $C(\mb\theta)$ is the expectation value of some 
physical observable that needs being minimized.

At the heart of many variational hybrid optimization processes, is the 
variational ansatz for the state $\ket{\psi(\mb\theta)}$
in \eqref{cost}. One of the most popular choices is to take such state 
as the output of a {\it parametric quantum circuit}
\begin{equation}
\ket{\psi(\mb\theta)} 
= e^{-i \theta_P \hat X_P}\cdots e^{-i\theta_1\hat  X_1} \ket{\psi_0}~,
	\label{ansatz}
\end{equation}
i.e.\;of a series of evolutions generated by different, and yet {\it fixed},
hamiltonian operators $\hat X_j$, for times 
$\theta_j$ representing the variational parameters.
The reason for this choice is that
parametric quantum circuits are implementable in NISQ 
devices \cite{preskill2018quantum} as long as $\hat X_j$ contains 1- and
2-local interactions only, i.e. when the gates 
$e^{-i\theta_j\hat  X_j}$ act non-trivially on at most two qubits. 
The state $\ket{\psi_0}$ is chosen among
states that are easy to prepare, and it is typically separable,
$\ket{\psi_0}\equiv \bigotimes_{j=1}^N |\psi_0^{(j)}\rangle$.

\begin{figure}[t]
	\centering
	\includegraphics[width=0.9\linewidth]{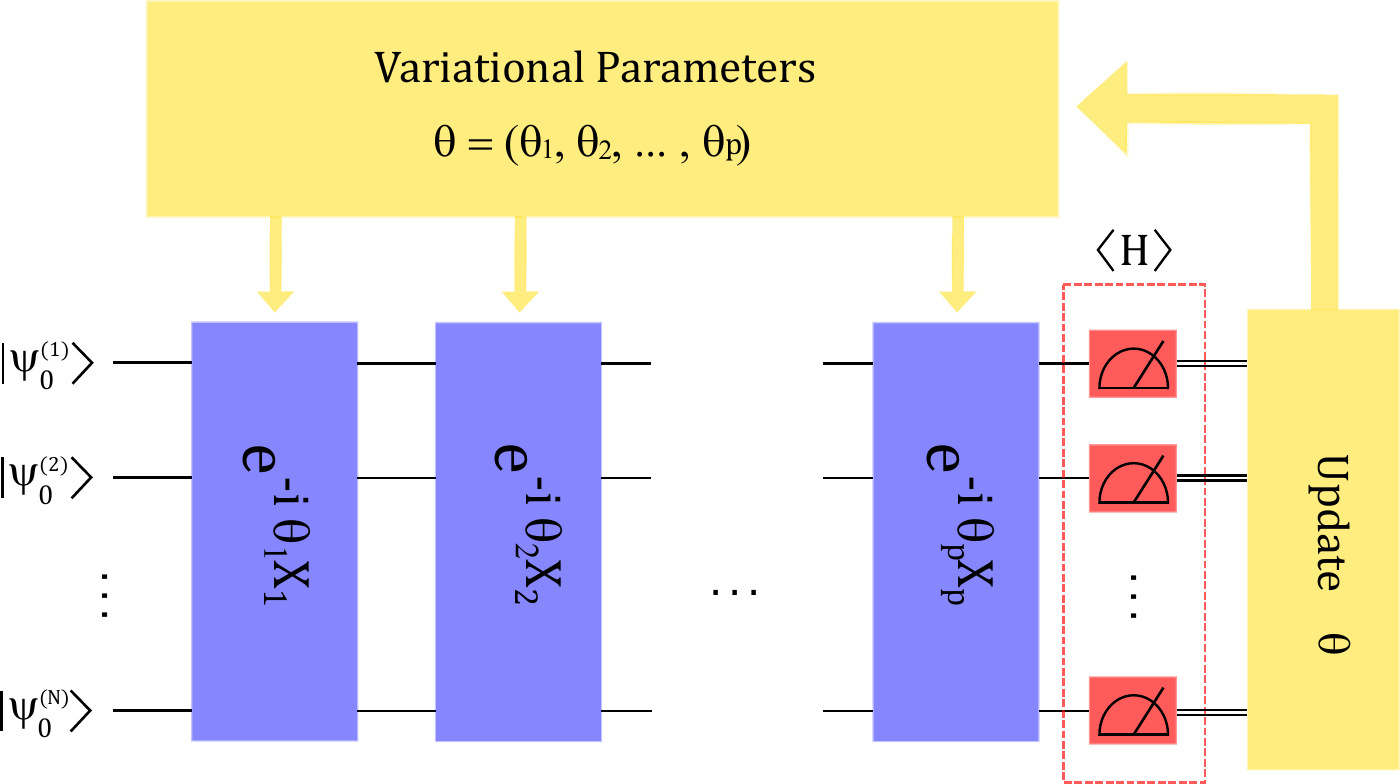}
	\caption{
		{ Variationl hybrid quantum-classical optimization.} A quantum computer is used to 
		prepare the variational state \eqref{ansatz} by sequentially applying some gates that 
		depend on parameters $\theta_j$, and then to measure the observable $\hat H$ 
		to estimate the cost \eqref{cost}.  A classical algorithm iteratively processes these
		outcomes and updates the parameters $\theta_j$ to iteratively minimize the cost 
		\eqref{cost}.
	}%
	\label{fig:scheme}
\end{figure}

Variational hybrid quantum-classical algorithms, schematically shown in
Fig.~\eqref{fig:scheme}, operate iteratively a quantum device and a 
classical processor.
At the $i$-th iteration, the quantum device 
generates the variational state 
$\ket{\psi(\mb\theta^{(i)})}=
e^{-i \theta^{(i)}_P \hat X_P}\cdots e^{-i\theta^{(i)}_1\hat  X_1} 
\ket{\psi(\mb\theta^{(i-1)})}$, and estimate the cost 
\eqref{cost},
and possibly its derivatives $\partial_{\theta_j} C$,
via quantum measurements 
\cite{mitarai2018quantum,schuld2019evaluating,harrow2019low} of the 
observable $\hat H$.
This is the computationally hardest part,
as it requires the manipulation of states that belong to Hilbert spaces 
whose dimension exponentially increases with the number of qubits $N$. 
Afterwards, a classical algorithm processes the estimated 
values of $C(\mb\theta^{(i)})$, or derivatives $\partial_{\theta_j} C$, 
and proposes new parameters $\mb\theta^{(i+1)}$ that 
are expected to flow towards the minimum of the cost function. 
Classical optimization, quantum evolution, and quantum measurements are 
thus performed iteratively till convergence.  
The advantage of this hybrid approach is that the quantum 
computer is always reset after each iteration so that the coherence 
times required are just those necessary to operate a circuit with depth 
$\mathcal O(P)$ and then perform a measurement.

The main difference with other common variational approaches used in 
quantum mechanics is that $C(\mb\theta)$, or derivatives 
$\partial_{\theta_j} C$, are estimated from measurement outcomes and, as 
such, are affected by uncertainty due to the probabilistic nature of 
quantum measurements, even in the noiseless case.

Having access to stochastic values of the cost function dramatically changes the convergence 
time \cite{bubeck2015convex}.
Algorithms for stochastic optimization are classified as zeroth-order, or derivative-free,
when only $C(\mb\theta)$ is measured, first-order
when it is possible to directly measure the derivatives w.r.t.\;$\theta_j$ 
of the cost function or, in general,
$k$th-order when also $k$th-order derivatives are available. 
It has been recently shown \cite{harrow2019low} that first-order methods can lead to substantially
faster convergence than zeroth-order methods. On the other hand, the
convergence time is not more strictly 
bounded when using higher-order derivatives, although some advantage may be observed in 
practical implementations. Motivated by that analysis, here we focus on the convergence of first-order 
methods using the framework of stochastic optimization. 

When dealing with stochastic optimization problems, where only
the stochastic outcomes $f(\mb\theta,y)$ are directly measurable by sampling 
different values of $y$ that are distributed according to a 
distribution $p(y|\mb\theta)$, the cost function is usually written 
\cite{kingma2014adam,bubeck2015convex,harrow2019low}
as $C(\mb\theta)=\mathbb E_{y \sim p(y|\mb\theta)} [f(\mb\theta,y)]$,
where $\mathbb E_{y \sim p(y|\mb\theta)} [f(\mb\theta,y)]$ is the 
expectation value $\sum_y p(y|\mb\theta)f(\mb\theta,y)$.
The cost function \eqref{cost} can be written in the above form by using 
the (possibly unknown) eigendecomposition 
of $\hat H\equiv \sum_y E_y \ket y \!\bra y$:
the measurement outcomes $y$ are distributed with probability 
$p(y|\mb\theta) = \bra y 
\hat \rho(\mb\theta) \ket y $,
where  $\hat \rho(\mb\theta) = \ket{\psi(\mb\theta)}\!\bra{\psi(\mb\theta)}$,
and $f(\mb\theta,y)=E_y$ is the associated cost, which is 
independent of $\mb\theta$.  

When the eigendecomposition 
of $\hat H$ is not known, one can still get $C(\mb\theta)$ from Pauli measurements,
namely by decomposing $\hat H$ as $\hat H= \sum_{\mu=1}^L h_\mu \hat \sigma_\mu$ 
where each $\hat \sigma_\mu$
is a tensor product of Pauli matrices and $h_\mu$ the corresponding coefficient,
and then by independently estimating each
$\langle\psi(\mb\theta) |\hat \sigma_\mu|\psi(\mb\theta)\rangle$. 
Note that many $\hat \sigma_\mu$ typically commute with each other, so the required 
number of independent measurements can be smaller than $L$.

Suppose now that $\mb\nabla C(\mb\theta)=\mathbb E_{z\sim q(z|\mb\theta)} [\mb g(\mb\theta,z)]$, 
with $\nabla_j\eqdef \frac{\partial}{\partial \theta_j}$,
i.e. that the gradient of $C$ can be written as an expectation of a 
vector-valued function $\mb g(\mb\theta,z)$ over some stochastic outcomes $z$,
distributed with a probability distribution $q$, possibly different from $p$. 
The simplest first-order method for stochastic optimization 
is stochastic gradient descent (SGD) that, intuitively, acts as a gradient 
descent algorithm where $\mb\nabla C$ is substituted 
with an unbiased estimate $\mb g$. If the parameters are updated 
at each iteration $i$
as $\mb\theta^{(i+1)} = \mb\theta^{(i)} - \alpha_i \mb g(\mb\theta^{(i)})$ then,
after $I$ iterations, the algorithm converges 
\cite{bubeck2015convex,harrow2019low,ruppert1988efficient,polyak1992acceleration} 
to a local optimum $\mb\theta^{\rm opt}$ with 
\begin{equation}
	\mathbb E[C(\mb\theta^{[1:I]})] - C(\mb\theta^{\rm opt}) \le R \frac{G}{\sqrt{I}}~.
	\label{bound}
\end{equation}
The left-hand side of \eqref{bound},  where
$\mb\theta^{[1:I]} = \frac1I\sum_{i=1}^I \mb\theta^{(i)}$,
formally defines what we call ``accuracy'' in this work;
in the right-hand side of the inequality
$R$ is a constant that depends on the function and on the parameter space,
while $G$ is an upper bound on the norm of the gradient estimate, 
$\mathbb E[\|\mb g(\mb\theta)\|_2^2] \le G^2$.
Such rate is achieved with $\alpha_i \equiv \alpha = R I^{-1/2}/G$. 
The inequality~\eqref{bound} means that a larger gradient variance implies slower convergence.
Note that, due to the stochastic nature of  $\mb g$, even the parameters
$\mb\theta^{(i)}$ are stochastic.
On the other hand, Eq.~\eqref{bound} shows that $\mb\theta^{[1:I]}$ is a {\it good estimator} 
of the optimal value $\mb\theta^{\rm opt}$ in the limit of many iterations $I$, and
an arbitrarily small 
error $\epsilon \propto G/\sqrt I$ may be achieved. 
In other algorithms \cite{bubeck2015convex,kingma2014adam,harrow2019low},
the convergence depends on the bound
$\mathbb E[\|\mb g(\mb\theta)\|_\infty^2] \le G_\infty^2$, obtained with a different norm. Since
norm inequalities imply $G\leq \sqrt{P} G_\infty$, we can always focus on $G_\infty$. 
Although different algorithms may have different convergence times, for instance with adaptive 
$\alpha_i$
and other definitions of $\mb\theta^{[1:I]}$, most upper bounds have 
a form similar to \eqref{bound}. Faster convergence, $\epsilon \approx G^2/I$,
can be obtained when $C(\mb\theta)$ satisfies extra properties \cite{bubeck2015convex,harrow2019low}, 
such as strong convexity, with a slightly different definition of $\mb\theta^{[1:I]}$. 
The bound \eqref{bound} assumes that the parameters are updated after each
query, namely after a single measurement outcome $\mb g$. An alternative is 
{\it mini-batch learning} \cite{bubeck2015convex}, 
where $M>1$ queries are used to better estimate the gradient. 
Although this yields a less-noisy gradient estimator, which for instance 
provides better numerical results in training quantum dynamical systems 
\cite{banchi2016quantum,innocenti2019supervised}, 
the theoretical worst-case convergence rate is similar to \eqref{bound}. Indeed, a bound like 
\eqref{bound} can be written with $I=M N_{\rm iter}$, with $N_{\rm iter}$ the number of iterations 
and $I$ the total number of measurements.


\section{Hybrid optimization with noisy operations}\label{s:bounds}
Due to the unavoidable errors in their operation, NISQ devices cannot exactly
prepare the ideal
variational state \eqref{ansatz}, which must hence be substituted with $\hat
\rho(\mb\theta) = \mathcal E(\mb\theta)[\hat \rho_0]$, where $\hat \rho_0$ is
the noisy version of $\ket{\psi_0}$ and
$\mathcal E(\mb\theta)$ the noisy dynamical map.
Although most of our theoretical bounds hold for more complex noise models,
for the sake of simplicity in the following we use the decomposition
\begin{equation}
        \hat \rho(\mb\theta) = \mathcal E^{\theta_P}_P\circ \cdots \circ \mathcal
E^{\theta_1}_1[\hat \rho_0]~,
        \label{noisyansatz}
\end{equation}
where $\circ$ indicates composition and $\mathcal E_j^{\theta_j}$ is the
noisy version of the ideal parametric unitary channel
$\mathcal U_j^{\theta_j}[\hat \rho] = e^{-i\theta_j\hat  X_j}\hat \rho
e^{i\theta_j \hat X_j}$
implemented by the $j$-th parametric gate of the NISQ device.
In what follows,
$C_{\rm min} := \min_{\psi}\langle\psi|H|\psi\rangle$ is the exact minimum of the cost function. 
Since $\hat \rho(\mb\theta)$ is a mixed state, 
the minimization of the cost function $C_{\rm noisy}(\mb\theta)\eqdef \Tr[\hat \rho(\mb\theta)\hat H]$ 
only provides an approximation 
to the minimum $C(\mb\theta^{\rm opt})$ that can be obtained in the noiseless case. 
Although not explicitly considered in this paper, 
our analysis may be straightforwardly extended to some cases where even 
the ideal target is a mixed state, 
e.g. the non-equilibrium steady state of a noisy evolution 
\cite{yoshioka2019variational,carollo2019quantumness,banchi2014quantum}.

The convergence rate of stochastic optimization towards the noisy minimum 
$C_{\rm noisy}(\mb\vartheta^{\rm opt.})$, 
with optimal parameters $\mb\vartheta^{\rm opt}$,
can be bounded as in Eq.~\eqref{bound}.
Considering both the error due to the finite number of iterations and the error due to 
the difference between $C(\mb\theta^{\rm opt})$ and 
$C_{\rm noisy}(\mb\vartheta^{\rm opt})$ we may write 
\begin{equation}
	C_{\rm noisy}(\mb\theta^{[1:I]})-C(\mb\theta^{\rm opt}) \le 
	{\rm Err}(\mb\theta^{\rm opt},\mb\vartheta^{\rm opt}) 
	+ R \frac{G^{\rm noisy}}{\sqrt I}~,
	\label{noisybound}
\end{equation}
where we define 
$	{\rm Err}(\mb\theta,\mb\vartheta) $
as the difference between the noisy and noiseless costs, 
namely 
\begin{equation}
	{\rm Err}(\mb\theta,\mb\vartheta) 
	\eqdef
	C_{\rm noisy}(\mb\vartheta)-C(\mb\theta)~.
	\label{err1}
\end{equation}
For some noise models, it can be shown that $\mb\vartheta^{\rm opt}=\mb\theta^{\rm opt}$,
namely that the optimal parameters in the noiseless and noisy case are the same \cite{sharma2019noise}. 
The inequality \eqref{noisybound} shows a simple and yet important aspect:
after a fixed number of iterations $I$, 
our best approximation to the noiseless variational minimum has an error that is given by two 
different terms. The first one follows from the difference between the 
noiseless and noisy case,
while the second one depends on the 
gradient estimator and always decreases for increasing $I$. 
To simplify our discussion and provide a worst-case scenario,
we assume that we know how to choose an ideal variational ansatz 
\eqref{ansatz} that provides  $C_{\rm min} = C(\mb\theta^{\rm opt})$, and consequently ensures 
${\rm Err}(\mb\vartheta^{\rm opt},\mb\theta^{\rm opt}) \geq 0$.
This is typically not the case, as variational ansatze are normally chosen as simple circuits that 
are easy to implement on a NISQ device, for which one might get a 
negative 
${\rm Err}(\mb\vartheta^{\rm opt},\mb\theta^{\rm opt})$.
The worst-case error coming from the first term in the r.h.s.\;of \eqref{noisybound} 
can be bounded by 
adapting the ``peeling'' technique from \cite{pirandola2017fundamental,pirandola2018theory}.
Indeed, we show in the Appendix~\ref{a:bound1} that
$ {\rm Err}(\mb\theta,\mb\vartheta) \leq P \|\hat H\|_\infty  
\max_k \|\mathcal E_k^{\vartheta_k}-\mathcal U_k^{\theta_k}\|_\diamond$ 
so the error increases at most linearly with the depth $P$ and depends on the maximum distance,
as measured by the diamond norm \cite{kitaev1997quantum,watrous2018theory}, between the ideal gates and their noisy implementations -- see also Eq.~\eqref{e:diamond}. 
An alternative inequality	${\rm Err}(\mb\theta,\mb\vartheta) \leq
2\|\hat H\|_\infty \;  \sqrt{1-\bra{\psi(\mb\theta)}
\hat \rho(\mb\vartheta)\ket{\psi(\mb\theta)}}$ shows that the first error term 
is bounded by the fidelity between the optimal pure state and its noisy version.

We now focus on $G^{\rm noisy}$ in \eqref{noisybound}, 
which depends on the procedure used to estimate the gradient from 
quantum measurements. 
The measurement of an observable with associated operator $\hat g_j$ 
provides an unbiased estimator of the gradient 
if $\nabla_j C = \Tr[\hat \rho\hat g_j]$ for each $j$. 
In this sense, we refer to the observables $\hat g_j$ 
as {\it estimators} of the gradient.
In the noiseless case different estimators have been proposed
\cite{mcclean2018barren,mitarai2018quantum,schuld2019evaluating,harrow2019low},
based on either the Hadamard test or the so-called parameter-shift rule. 
However, those estimators may result biased if only noisy gates are 
available:
therefore, a rigorous generalization to the noisy regime 
is still lacking. 
The convergence of SGD with biased gradient estimators is not much understood, 
aside from specific algorithms such as simultaneous perturbation stochastic
approximation (SPSA) \cite{spall1992multivariate}, where the bias 
can be controlled. 
In order to define an unbiased estimator in the general case we use the geometry 
of quantum states, from which we know that any derivative 
can be written as \cite{paris2009quantum,bengtsson2017geometry} 
\begin{equation}
	\nabla_j \hat \rho = \frac{\hat L_j \hat \rho + \hat \rho \hat L_j}2~,
	\label{SLD}
\end{equation}
where the operator $\hat L_j$ is called the symmetric logarithmic derivative (SLD). 
The gradient of the cost $C(\mb\theta)=\Tr[\hat \rho(\mb\theta)\hat H]$ 
can hence be obtained by measuring observables with associated operators
\begin{equation}
	\hat g_j(\mb\theta) = \frac{\hat L_j(\mb\theta) \hat H+\hat H \hat L_j(\mb\theta)}2 + \lambda_j \hat L_j(\mb\theta)~,
	\label{grad}
\end{equation} 
for any $\lambda_j$.  
The freedom in choosing $\lambda_j$ follows from Eq.~\eqref{SLD}, as
$\Tr[\hat
L_j\hat \rho] = \Tr [\nabla_j \hat \rho] 
= \nabla_j \Tr[\hat \rho]=0$ implies that the expectation value 
$\nabla_j C=\Tr[\hat g_j \hat \rho]$ 
is independent of $\lambda_j$. Therefore, the free parameters $\lambda_j$ are analogous to the 
so-called baselines, commonly employed in reinforcement learning for variance
reduction \cite{deisenroth2013survey}. The optimal $\lambda_j$s are discussed in 
Appendix~\ref{a:baselines}.
The measurement of the gradient operators provides
stochastic outcomes $g^{\rm SLD}_j(\mb\theta,\gamma)$ with probabilities
$\langle{g_{\gamma,j}}\vert\hat \rho\vert{g_{\gamma,j}}\rangle$, 
where we used the eigendecomposition
$\hat g_j {=} \sum_{\gamma} g^{\rm SLD}_j(\mb\theta,\gamma)
|{g_{\gamma,j}}\rangle\!\langle{g_{\gamma,j}}|$. 
For pure states, the SLD operator has a simple form 
$\hat L_j = \ket{\psi(\mb\theta)}\langle{\nabla_j\psi(\mb\theta)}\vert$ and the above 
estimation strategy becomes equivalent to others, proposed in the 
literature 
\cite{mcclean2018barren,mitarai2018quantum,schuld2019evaluating,harrow2019low},
which can be explicitly measured using a generalization of the Hadamard test \cite{harrow2019low}.

An alternative estimator can be obtained using 
the log-derivative (LD) trick 
\cite{glynn1990likelihood}, 
also called ``reinforce'' in the machine learning literature \cite{williams1992simple},
which consists in writing the gradient of the cost function 
$\nabla_j C = \sum_y E_y \nabla_j p(y|\mb\theta)$ 
as en expectation value of 
$g^{\rm LD}_j(\mb\theta,y) = E_y \nabla_j \log p(y|\mb\theta)$ over the original distribution  
$p(y|\mb\theta) = \bra y \hat \rho(\mb\theta)\ket y$ where $\hat H=\sum_y E_y \ket y\!\bra y$. 

In Appendix~\ref{a:bound2}, 
we show that all different estimators for the gradient 
satisfy the upper bound 
\begin{equation}
	G^{\rm noisy} \leq \sqrt{P} G^{\rm noisy}_\infty \leq 
 \sqrt{P} \|\hat H\|_\infty \max_{j,\mb\theta} \sqrt{{\rm QFI}_j(\mb\theta)}~,
\label{QFIbound}
\end{equation}
where QFI is the Quantum Fisher Information 
\begin{equation}
	{\rm QFI}_j(\mb\theta) = \Tr[\hat \rho(\mb\theta) \hat L_j(\mb\theta)^2]~,
	\label{QFI}
\end{equation}
a central quantity in quantum metrology \cite{paris2009quantum} 
that is also relevant for studying 
quantum phase transitions \cite{zanardi2007information,venuti2007quantum,banchi2014quantum}. 

Before proceeding further, let us briefly comment upon the emergence of 
the QFI in this context.
There are two slightly different way of understanding the QFI.
First, there is the QFI of a state $\hat \rho$ {\it with respect to a 
certain observable $\hat X$}: this is $F_{Q}[\hat\rho,\hat X]
:=2\sum_{mn}\frac{(\rho_m-\rho_n)^2}{(\rho_m+\rho_n)}
|\bra{m}\hat X\ket{n}|^2$, where $\hat\rho=\sum_n \rho_n \ket n\!\bra n$ is 
the eigendecomposition of $\hat\rho$.
When $\hat\rho$ is a pure state, the QFI is nothing 
but twice the expectation value of $\hat X$ on such state.
On the other hand, if it exist a reference state $\hat\rho_0$ and a real 
parameter $\theta$ (that we here take as one-dimensional for the sake of 
clarity) such that $\hat\rho=e^{-i\theta\hat X}\hat\rho_0e^{i\theta\hat X}\:=\hat\rho(\theta)$, then 
it is $F_Q[\hat\rho,\hat X]=F_Q[\hat\rho(\theta)]={\rm Tr}[\hat\rho(\theta)\hat 
L^2]$, where $\hat L$ is the LSD we have encounterd above, whose relation 
with the observable $\hat X$ is via 
$i[\hat\rho,\hat X]=\frac{1}{2}(\hat L\hat\rho+\hat\rho\hat L)$.
The expression $F_Q[\hat\rho(\theta)]$ is common in quantum 
metrology, where it quantifies the sensitivity of the parametric state 
$\hat\rho(\theta)$ to the variations of $\theta$ itself. To this respect
notice that if $\theta$ represents the classical quantity to be 
measured via a quantum-metrology protocol, a higher QFI guarantees a 
more precise result. The above comments are easily generalized 
to the case of a multidimensional parameter $\mb\theta$ as the one used 
elsewhere in this work.

\begin{figure}[t]
	\centering
	\includegraphics[width=0.9\linewidth]{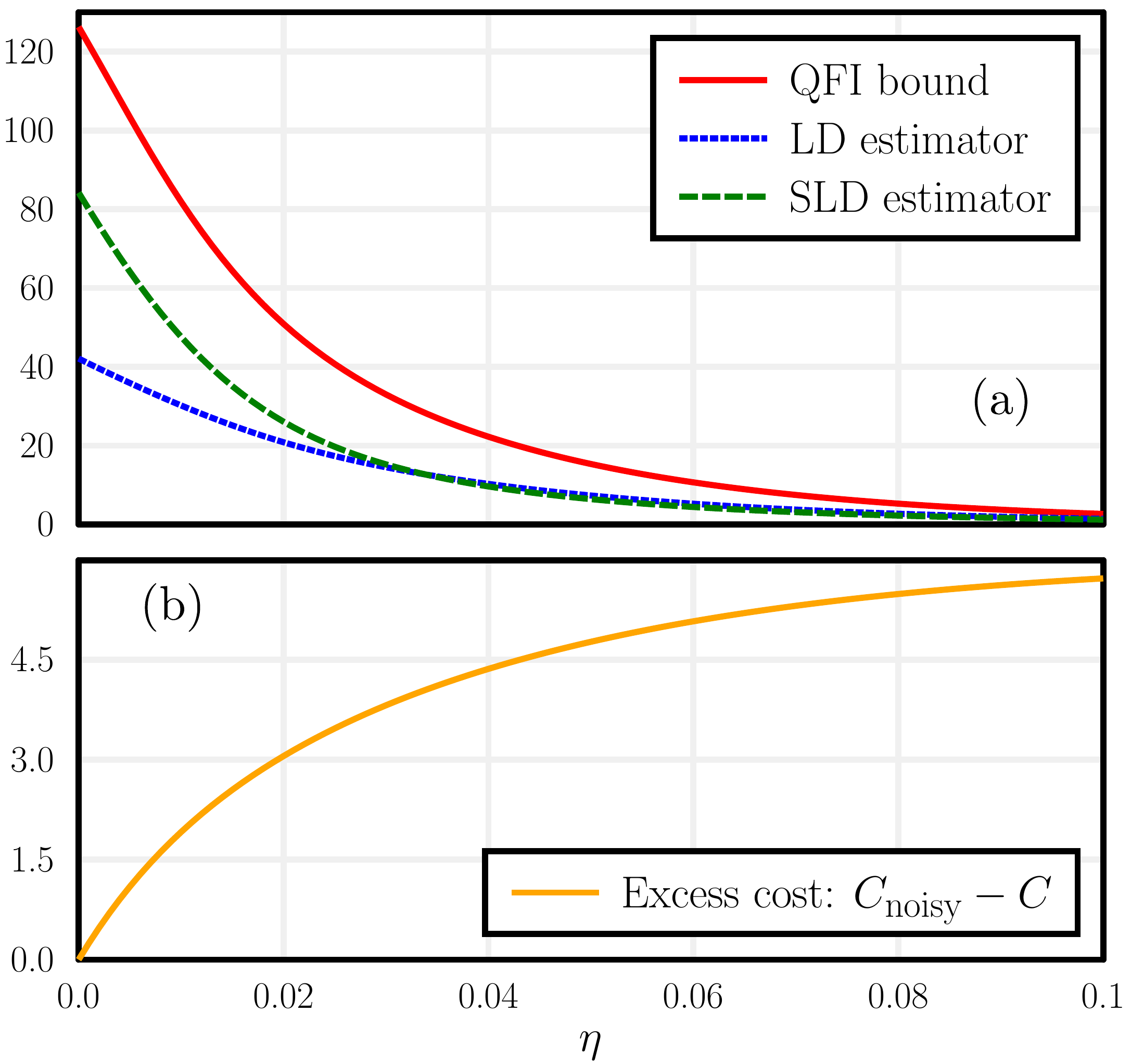}
	\caption{{ Opposite behaviour of the two sources of error.} (a) the square
		root of second statistical
		moment of the gradient estimator 
		$\max_{\mb\theta}\sqrt{\mathbb E[\|\mb
		g(\mb\theta)\|_2^2]} \leq G_{\rm noisy}$  and (b) the excess cost
		${\rm Err}(\mb\theta^{\rm opt},\mb\vartheta^{\rm opt}) $,
		as a function of the depolarising noise strength $\eta$ (see Sec.~\ref{s:numerics} for the definition of $\eta$). The variational circuit 
		corresponds to a QAOA for a ring of $N=6$ qubits with $20$ variational
		parameters. Different gradient estimators are considered: the one based on
		the log-derivative trick (LD) and the one based on the symmetric 
		logarithmic derivative (SLD). Those are plotted against the upper bound \eqref{QFIbound} 
		based on the Quantum Fisher Information (QFI).
	}
	\label{fig:bounds}
\end{figure}

\begin{figure}[]
	\centering
	\includegraphics[width=0.99\linewidth]{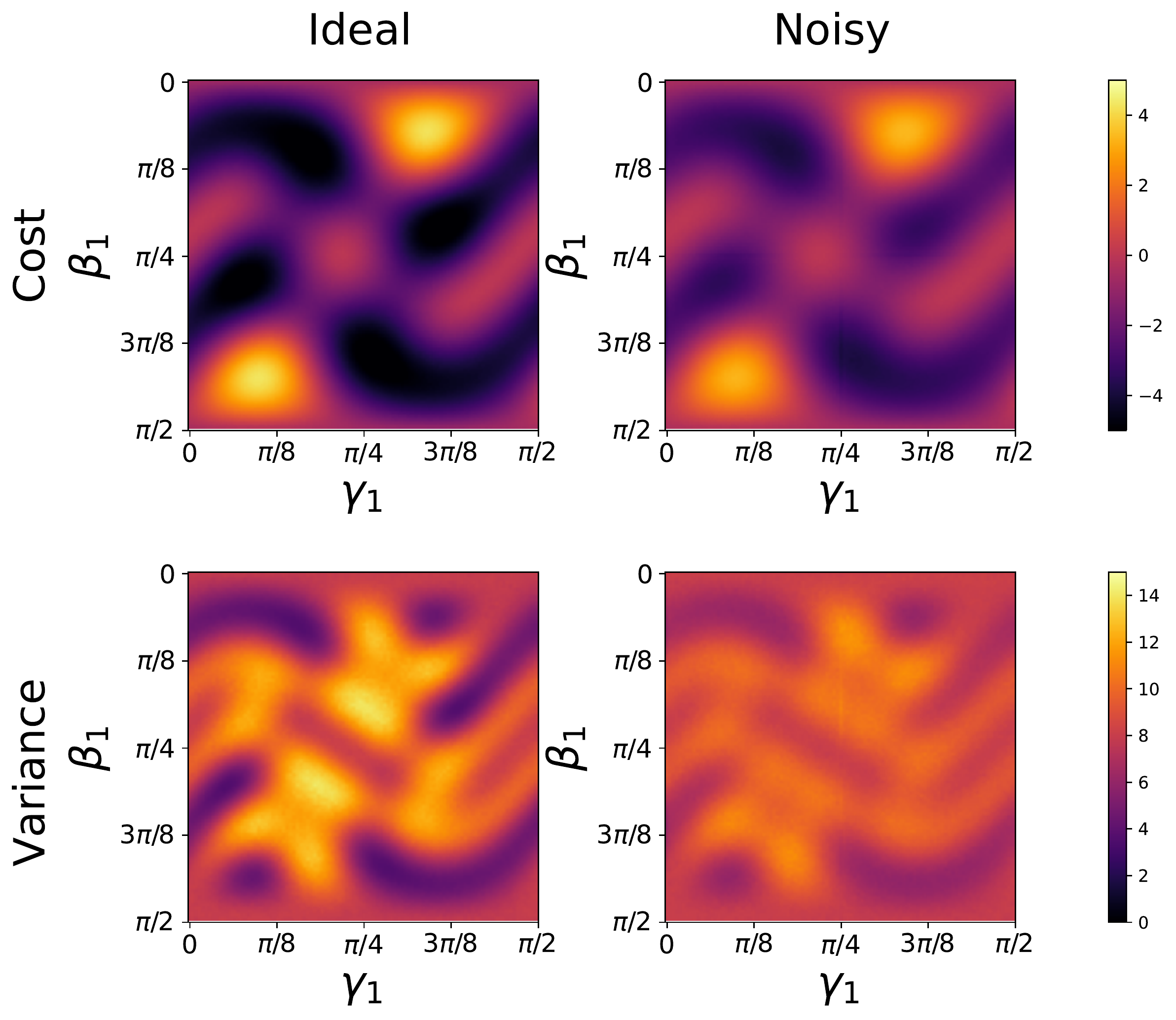}
	\caption{ Typical form of the cost-function landscape in the noiseless (left) to the noisy (right)
		regimes as a function of two parameters, $\gamma_1$ and $\beta_1$ --
		see Sec.~\ref{s:numerics} for the model description. 
		Costs $\langle \hat H\rangle$ and variances $\langle \hat H^2\rangle -\langle \hat H\rangle^2$ are respectively 
		shown in the top and bottom panels. 
	}%
	\label{fig:landscape}
\end{figure}

Let us now get back to the bound \eqref{QFIbound} to underline 
a very important aspect: while the first term in the r.h.s.\;of 
\eqref{noisybound} 
increases as a function of the noise strength, the second one can decrease. 
Indeed, it is known that noise is normally detrimental for metrology, 
as it can reduce the QFI from $\mathcal O(N^2)$ (Heisenberg limit) to $\mathcal O(N)$ (standard 
quantum limit) \cite{escher2011general,giovannetti2011advances}. 
Our analysis thus shows that the convergence accuracy, as defined by 
the l.h.s.\;of \eqref{noisybound}, is bounded by the sum of
two terms that typically display opposite behaviours as a function 
of the noise strength, with the first one increasing and the 
second one decreasing,
as shown in Fig.~\ref{fig:bounds} for the specific example 
that will be described in the following section.
From the physical point of view, this is due to how the cost-function landscape is 
affected by noise. As shown in Fig.~\ref{fig:landscape} for strong noise the
landscape is ``flattened'', 
and hence the minimum cost is higher, but so does the variance, 
and hence the maximum variance is lower. Therefore, in the noisy 
regime the number of operations needed in order to correctly find the 
minimum may be reduced. 
Our analysis of Eq.~\eqref{noisybound} quantifies this intuition: 
when the number of iterations $I$ is small, we may observe that noise 
is actually beneficial. However, due to measurement uncertainty, the number 
of iterations required for convergence is typically large and 
 $I\gg \sqrt PR^2 {\rm QFI}$.  In this regime
noise is always detrimental, as observed in some numerical experiments
\cite{xue2019effects,alam2019analysis}.

The bound \eqref{noisybound}, together with the above analysis, shows that 
the convergence speed is mostly unaffected by noise, while the quality of the 
solution typically deteriorates due to the error ${\rm Err}(\mb\theta,\mb\vartheta)$,
which still increases at most linearly with the depth $P$ of the circuit. 
Therefore, we may expect that hybrid variational optimization is robust against 
noise. In the next section we numerically study such robustness for a relevant 
optimization problem.

\section{Numerical examples }\label{s:numerics}
QAOA \cite{farhi2014quantum} 
is a specific ansatz for variational hybrid optimization which consists in 
the repetition of two types of parametric quantum evolutions generated by 
two different non-commuting Hamiltonians, typically called $\hat H_\gamma$ and 
$\hat H_\beta$. Here
$\hat H_\gamma \equiv \hat H$ is 
equal to the cost operator appearing in Eq.~\eqref{cost} and is a function 
of the Pauli $\hat \sigma^z_l$ operators only, where the indices $l=1,\dots,N$ refer
to the different qubits. In the {\it computational basis} 
defined by the eigenstates $\{\ket0,\ket1\}$ of $\hat \sigma_l^z$, $H$ is diagonal.
The other Hamiltonian is fixed as 
$\hat H_\beta = -\sum_l \hat \sigma^x_l$, where $\hat \sigma_l^x$ are other Pauli 
operators, which are not diagonal in the computational basis.
The QAOA evolution can be written as in Eq.~\eqref{ansatz} with 
sequential applications of $\hat H_\gamma$ and $\hat H_\beta$ 
\begin{equation}
	\ket{\psi(\mb\gamma,\mb\beta)} = 
	e^{-i\beta_{\cal P} \hat H_\beta} 
	e^{-i\gamma_{\cal P} \hat H_\gamma}  \cdots
	e^{-i\beta_1 \hat H_\beta} 
	e^{-i\gamma_1\hat  H_\gamma} 
	\ket{+}^{\otimes N}~.
	\label{qaoa}
\end{equation}
The parameters are then split as $\mb\theta=(\mb\gamma,\mb\beta)$ and the total depth 
of the circuit is $P=2{\cal P}$. The initial state $\ket{\psi_0}=\ket{+}^{\otimes N}$,
where $\ket{+}=(\ket0+\ket 1)/\sqrt 2$, 
is the ground state of $\hat H_\beta$. 
QAOA is a universal model for quantum computation 
\cite{lloyd2018quantum,morales2019universality}, 
meaning that, with specific choices of $\hat H_\gamma$,
any state can be arbitrarily well approximated 
by $\ket{\psi(\mb\gamma,\mb\beta)}$ with suitable parameters $\gamma_j$, $\beta_j$
and ${\cal P}\to\infty$.
For the specific choice $\gamma_j\propto j/{\cal P}$ and $\beta_j\propto (1-j/{\cal P})$,
Eq.~\eqref{qaoa} can be interpreted as a discretization  of an adiabatic evolution 
\cite{barends2016digitized,farhi2014quantum} and
QAOA is guaranteed to perform well 
for large enough ${\cal P}$.
The QFI can be large when the adiabatic evolution crosses a dynamical phase transition 
\cite{zanardi2007information,venuti2007quantum,banchi2014quantum} and the 
error from $G_{\rm noisy}$ in \eqref{noisybound} 
can be significant when the Hamiltonian $\beta \hat H_\beta + \gamma \hat H_\gamma$ 
displays a quantum phase transition for some choices of $(\beta,\gamma)$. 
One such example is the Ising ring \cite{lieb1961two} studied below, 
where $\hat H_\beta$ models the global transverse field. In such model,
the QFI scales as $\mathcal O(N^2)$ in the translational invariant case, 
while for steady states of more complex noisy evolution 
it can be as large as $\mathcal O(N^6)$ \cite{banchi2014quantum}. 

Here we study QAOA applied to a translational invariant antiferromagnetic ring with 
$\hat H_\gamma=\sum_{l=1}^N \hat \sigma_l^z \hat \sigma_{l+1}^z$ and
periodic boundary conditions $\hat \sigma^z_{N+1}\equiv \hat \sigma^z_1$. QAOA with 
this model has been 
studied in \cite{wang2018quantum,mbeng2019quantum}, using the exact mapping 
to a free-fermion model. In particular, it has been proven \cite{wang2018quantum}
that the ground state 
can be exactly expressed with the QAOA ansatz \eqref{qaoa} as long as ${\cal P}\geq N/2$.
The effect of noise in an overparameterized QAOA is shown in Fig.~\ref{fig:bounds},
where we consider the effect of a local depolarising error, as in  \eqref{noisyansatz} 
with $\mathcal E_j^{\theta_j}[\hat\rho] =
\mathcal D[e^{-i\theta_j \hat X_j}\hat\rho e^{i\theta_j\hat X_j}]$, 
$\mathcal D = \bigotimes_{l=1}^N\mathcal D_l$ and 
$\mathcal D_l(\rho) = (1-\eta)\hat\rho +\eta \hat\sigma_l^z\hat\rho\hat\sigma_l^z$.
All bounds are computed by numerically finding the operators $\hat L_j$
from Eq.~\eqref{SLD}.  In Fig.~\ref{fig:bounds} we 
see that our theory predicts a decreasing $G_{\rm noisy}$ in \eqref{noisybound} 
as a function of $\eta$. 
In Appendix~\ref{a:fluct} 
we also study a different noise model, where the NISQ 
computer implements noisy yet unitary gates $e^{-i(\theta_j+\eta\epsilon_j)X_j}$ 
where $\epsilon_j\sim \mathcal N(0,1)$ 
is a Gaussian random variable. We found that also with this noise, 
the error terms display the same behaviour shown in Fig.~\ref{fig:bounds}.

\begin{figure}[ht!]
	\centering
	\includegraphics[width=\linewidth]{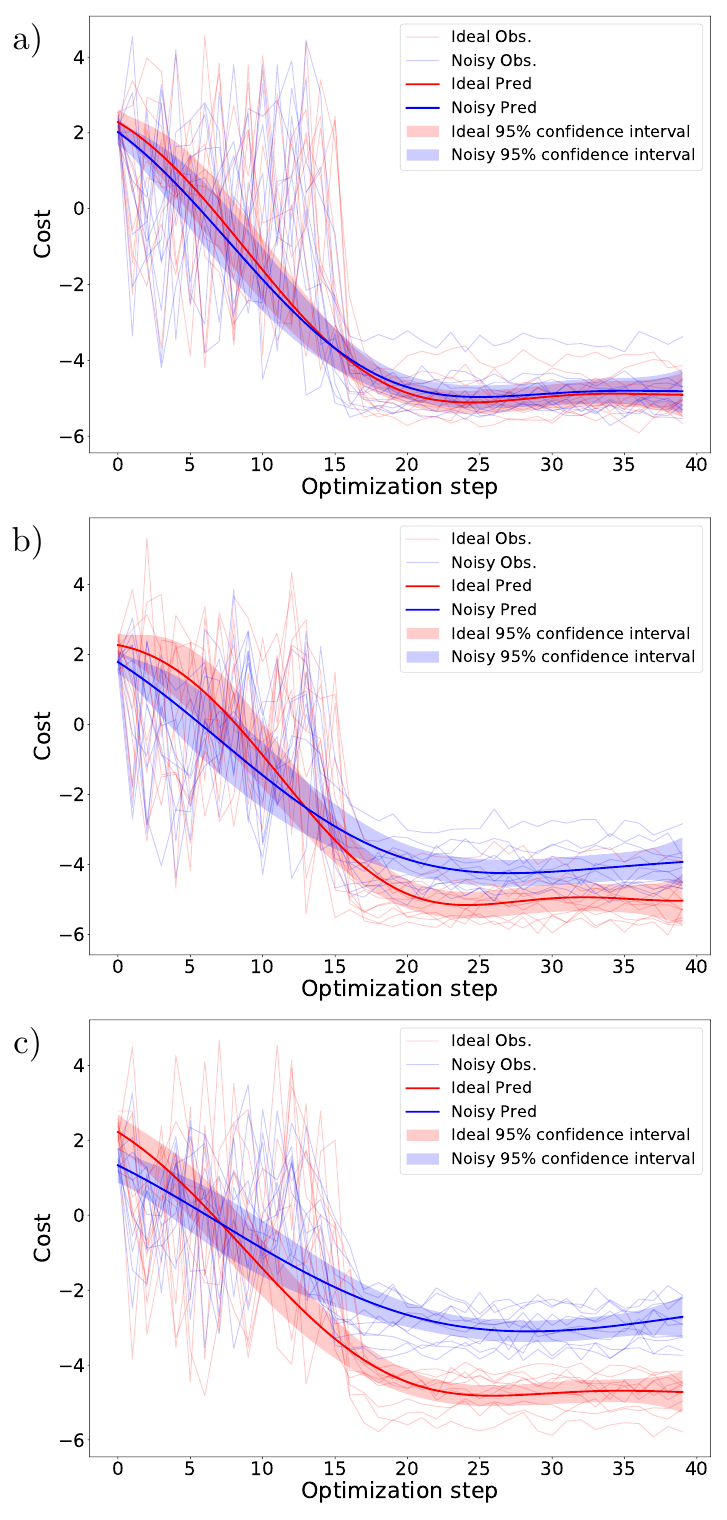}
	\caption{ {Numerical results on the convergence of QAOA.} 
		We use $N=8$ and ${\cal P}=3$ in all panels a), b) and c). 
		Red and blue curves, respectively, represent noiseless and noisy evolution.  
		The noise parameters are taken from typical values of 
		IBM quantum processors (see Appendix~\ref{a:noise}), yet increased by a 
		factor  2 (panel~a), a factor 4 (panel~b) and a factor 10 (panel~c). 
                The convergence of the cost function \eqref{cost} is shown 
		during the optimization procedure;  faded thin lines are ten trials of
		the same realization of QAOA,
		while solid lines show the empirical average evolution from the ten trials.
		For more information, see the main text.
	}%
	\label{fig:hist}
\end{figure}

We test our theoretical predictions using the Qiskit framework \cite{Qiskit} to simulate QAOA on a physical hardware by numerical experiments. 
In these simulations, the error model 
consists of 
single- and two-qubit gate errors, i.e.\;depolarizing error followed by a thermal relaxation error,
as discussed in Appendix~\ref{a:noise}. 
In Fig.~\ref{fig:hist} we show the optimization of the cost function 
\eqref{cost} using a simple gradient descend update, where each gradient value 
is estimated by using a quantum circuit followed by local measurements 
(see Appendix~\ref{a:gradient} for details); 
these operations are repeated 200 times to estimate a single value. 
Due to the stochastic measurement outcomes, each optimization takes a different 
stochastic route. 
Fig.~\ref{fig:hist} is divided into three panels, a), b) and c), differing from each other only for noise 
strength. In fact, in order to study the performance of QAOA in different noisy regimes, we progressively
increase noise strength in our simulations: in the results reported in Fig.~\ref{fig:hist} 
noise strength has been progressively increased by factor $f=2$ (panel~a), a factor $f=4$ (panel~b),
and a factor $f=10$ (panel~c), namely
gate errors, relaxation and decoherence rates are all increased by the factor $f$ 
compared to the typical values of noise parameters obtained from IBM quantum processors (see Appendix~\ref{a:noise}). 
Results obtained without increase, namely with $f=1$, are not reported, since they are basically indistinguishable 
from those obtained in the noiseless case. 
We expect that noise effects can become visible even with $f=1$ when the number of qubits $N$ and/or circuit depth $P$ are
sensibly larger than those employed in our simulations. 

The following description holds for all three individual panels regardless of the increase factor $f$ of noise strength:
the fade thin lines correspond to ten different
trials of optimization of the cost function in both noisy (blue) and noiseless (red) cases with
the same initial conditions. We observe that the cost, estimated alike the gradient repeating 200 times the
quantum circuit and measurements, initially highly fluctuates between
different runs, but then all runs converge towards the local minimum 
(since ${\cal P}<N/2$ the global minimum cannot be achieved). Higher accuracy in the
convergence may be obtained using more sophisticated first order methods 
\cite{kingma2014adam}. Optimizations starting from
different initial parameters display a similar behaviour (not shown). 
Using Gaussian process regression we have also computed average ``path-integral-like'' optimization curves,
which are shown in the same graph in Fig.~\ref{fig:hist}
with solid lines for both the noiseless and noisy cases. These lines show the empirical average evolution 
from the ten trials, 
and colored zones represent 95\% confidence intervals associated with such average.

We observe that for small noise strength, e.g. panel a) in Fig.~\ref{fig:hist},
the average optimization curves are basically indistinguishable
from the noisy ones, aside from finite-size effects due to the finite number of measurements. 
For moderate noise values, panel b) in Fig.~\ref{fig:hist}, as predicted by our theory, 
the final point in the optimization is slightly higher in the noisy regime, but the convergence speed is comparable.
Moreover, at the initial points, the average cost is smaller in the noisy case. 
Although this prediction may be due to the finite number of curves, it is consistent with our analysis from 
\eqref{noisybound}. 
For large noise strengths, panel c) in Fig.~\ref{fig:hist}, we observe that the final cost significantly deviates 
from the noiseless case, but the convergence speed is again comparable.

Overall our analysis shows that the convergence speed of hybrid variational optimization is 
not significantly affected by noise, while the quality of the final solution is.

\section{Discussion}\label{s:conclusion}

We have studied the convergence speed of variational hybrid quantum-classical
optimization algorithms, showing that the error after a finite number of steps 
can be upper-bounded by the sum of two terms:
the first one is the difference between the noisy and the noiseless
result, and typically increases for stronger noise;
the second term, though, is proportional to the square root of the
quantum Fisher information, that usually decreases with noise.
Due to the competition between these two terms, depending on the noise 
strength, different results can be observed. 
For small to intermediate noise we find that the convergence time 
is mostly unaffected by the presence of noise; 
This said, for a number of iterations such that the convergence 
time has not passed yet, the accuracy may actually be higher in some 
noisy regimes. On the other hand, if the convergence time is reached, 
the error term typically dominates, showing the foreseeable negative 
role of noise. 

Let us also comment upon the way QFI enters our results. 
From the formal viewpoint, we understand its occurrence as due to 
the use of stochastic optimization methods,
which involve the gradient of the cost function with respect to the 
variational parameters, and hence the operators 
$\hat{L}_j$ in \eqref{SLD}, and QFI via its definition \eqref{QFI}. 
However, the substantial reason for the QFI to appear in 
\eqref{QFIbound} in a somehow counterintuitive way (the lower the 
better) can be explained as follows.
In estimation theory a 
larger FI guarantees a better determination of the wanted parameter via 
the 
sampling of a function that depends on it; this is because a 
larger FI follows from larger local values of the derivatives, and hence 
a higher sensitivity of the overall estimation procedure. 
On the other hand, in the scheme to which we are referring
the role played by the parameter and the sampled function is reversed: 
we input different values of $\mb{\theta}$ aiming at exploring the 
$C(\mb{\theta})$-landscape, possibly locating its minimum; 
this exploration is more agile if the above landscape is more 
level, which corresponds to a lower FI. This general argument 
holds both in a classical and in a quantum setting, and we think it 
lies underneath the result Eq.~\eqref{QFIbound} in the following sense:
noise can help an algorithmic procedure to more easily explore 
the landscape of the cost function one wants to minimize, thus 
increasing, at least as far as its detrimental effect on the cost-function 
evaluation is not too strong, the overall efficiency of the 
optimization scheme.

We finally underline that QFI enters our analysis by only
providing a theoretical upper bound that never needs being
evaluated. In fact, should the QFI be
efficiently measurable, one could use more sophisticated stochastic algorithms, such as Amari's natural gradient 
\cite{amari1998natural}; this has been recently applied to noiseless 
parametric quantum circuits \cite{stokes2019quantum} based on the fact that, when
$C=-\log p(x,\mb\theta)$, the natural gradient is Fisher efficient, i.e. such that
the variance of the estimator $\mb\theta^{[1:I]}$ asimptotically meets the Cram\'er-Rao lower bound. 
However, such a result does not hold for more general cost functions like \eqref{cost}. 
Furthermore, no efficient method (e.g.\;poly$(N)$) for estimating the QFI 
from measurements is currently available in the noisy regime and, even if
it existed, estimating the QFI at each step would require further quantum measurements that would increase the query
complexity. In fact, understanding whether one can obtain Fisher efficient estimators of the optimal parameters is currently an open 
question.

\begin{acknowledgements}
 L.B. acknowledges support by the program ``Rita Levi Montalcini'' for young researchers. 
 S.P. acknowledges support by the
 QUARTET project funded by the European Union’s Horizon 2020 (Grant agreement No 862644).
 This work is
done in the framework of the Convenzione operativa between
the Institute for Complex Systems of the Consiglio Nazionale
delle Ricerche (Italy) and the Physics and Astronomy Department of the University of Florence.
\end{acknowledgements}

\appendix

\section{Bound on $ {\rm Err}(\mb\theta,\mb\vartheta) $ }\label{a:bound1}

Many of our results hold irrespective of assumption 
\eqref{noisyansatz}, and are valid for any error model 
\begin{equation}
	\hat \rho(\mb\theta) = \mathcal E(\theta_1,\dots,\theta_P)[\hat\rho_0]~.
	\label{asdgen}
\end{equation}
Indeed, all results that we derive in appendices \ref{a:baselines}, \ref{a:bound2} 
do not depend on the assumption \eqref{noisyansatz} and are valid for 
any noise model as in \eqref{asdgen}.
Here we show on the other hand that when the local error model
\eqref{noisyansatz} is assumed, then the error
${\rm Err}(\mb\theta,\mb\vartheta)$ grows at most linearly 
with the number of parameters. 
We study an upper bound to the first error in \eqref{noisybound},
which is clearly valid irrespective of the sign of ${\rm Err}(\mb\theta,\mb\vartheta)$ 
\begin{align}
	{\rm Err}(\mb\theta,\mb\vartheta) &\eqdef 
	\Tr[\hat H(\hat\rho(\mb\vartheta)-\ket{\psi(\mb\theta)}\!\bra{\psi(\mb\theta)})]
																 \nonumber\\ & \stackrel{(a)}{\leq}
	\|\hat H\|_\infty \; \| \hat\rho(\mb\vartheta)-\ket{\psi(\mb\theta)}\!\bra{\psi(\mb\theta)}\|_1
																 \nonumber\\ & \stackrel{(b)}{\leq}
  \|\hat H\|_\infty \; \| \mathcal E(\mb\vartheta)-\mathcal U(\mb\theta)\|_\diamond~,
	\label{a:in1}
\end{align}
where $\|\hat X\|_\infty$ is the maximum singular value of $\hat X$,
namely the maximum absolute value $|x_j|$ 
where $x_j$ are the eigenvalues of $\hat X$, $\|\hat X\|_1 = \Tr[\sqrt{\hat
X\hat X^\dagger}]$ is the trace norm,
and $\|\mathcal X\|_\diamond$ is the diamond norm for quantum channels
\cite{kitaev1997quantum,watrous2018theory}.
In the last line it is 
\begin{align}
	\mathcal U(\mb\theta) 
	&\eqdef
	\mathcal U_P^{\theta_P} \circ \cdots \circ \mathcal U_1^{\theta_1}~,
\end{align}
and
\begin{align*}
	\hat \rho(\mb\vartheta) &= \mathcal E(\mb\vartheta)[\ket{\psi_0}\!\bra{\psi_0}]~,
										 &
\ket{\psi(\mb\theta)}\!\bra{\psi(\mb\theta)} &= \mathcal U(\mb\theta)[\ket{\psi_0}\!\bra{\psi_0}]~,
\end{align*}
where for simplicity we have absorbed the noisy preparation of $\ket{\psi_0}$ into $\mathcal E_1$. 
To derive \eqref{a:in1},
in (a) we used the H\"older inequality and in (b) we used the distance induced by the 
diamond norm 
\begin{equation}
	\|\mathcal E-\mathcal U\|_\diamond = \max_\rho \|\mathcal I\otimes \mathcal E(\rho) - \mathcal I\otimes \mathcal U(\rho)\|_1~,
	\label{e:diamond}
\end{equation}
where $\mathcal I$ is the identity channel. 
We can now apply the ``peeling'' technique from \cite{pirandola2017fundamental,pirandola2018theory}
to bound the error in the diamond distance. To this aim, we now use the decomposition 
from Eq.~\eqref{noisyansatz} from the main text, and let 
$\delta_P = \|\mathcal E_{1:P}-\mathcal U_{1:P}\|_\diamond$,
where the $1{:}k$ refers to the composition of the first $k$ channels. Then, using the monotonicity 
of the diamond norm over CPTP maps and the triangle inequality, we may write
\begin{align*}
&	\delta_P = 
\\&
	\| \mathcal E_P\circ \mathcal E_{1:P-1} -
	 \mathcal E_P\circ \mathcal U_{1:P-1} +
	 \mathcal E_P\circ \mathcal U_{1:P-1} -
	 \mathcal U_P\circ \mathcal U_{1:P-1}\|_\diamond
				 \\&\leq
	\| \mathcal E_P\circ \mathcal E_{1:P-1} -
	 \mathcal E_P\circ \mathcal U_{1:P-1}\|_\diamond 
				 +\\&\quad +\|
	 \mathcal E_P\circ \mathcal U_{1:P-1} -
	 \mathcal U_P\circ \mathcal U_{1:P-1}\|_\diamond
					\\& \leq \delta_{P-1} + \|\mathcal E_P-\mathcal U_P\|_\diamond~.
\end{align*}
Iteratively applying the above inequality one gets
\begin{equation}
	\delta_{P} \leq \sum_{k=1}^P \|\mathcal E_k-\mathcal U_k\|_\diamond
	\leq P \max_k \|\mathcal E_k-\mathcal U_k\|_\diamond~.
	\label{asd}
\end{equation}
Combining \eqref{asd} and \eqref{a:in1} we find that the 
error increases at most linearly with $P$, according to 
\begin{align}
	{\rm Err}(\mb\theta,\mb\vartheta) &
	\leq P \|\hat H\|_\infty \max_k \|\mathcal E_k-\mathcal U_k\|_\diamond~.
\end{align}

An alternative bound can be obtained from \eqref{a:in1} 
via the Fuchs-van de Graaf inequality 
\cite{fuchs1999cryptographic}
\begin{align}
	{\rm Err}(\mb\theta,\mb\vartheta) \leq
		2\|\hat H\|_\infty \;  \sqrt{1-\bra{\psi(\mb\theta)}
		\hat \rho(\mb\vartheta)\ket{\psi(\mb\theta)}}~.
\end{align}

\section{Optimal baselines}\label{a:baselines}
We discuss the role of the free parameters $\lambda_j$, dubbed ``baselines'', in the optimization. 
In principle, such parameters should be chosen
to minimize 
$ \mathbb E [g_j^2 ] $.
We may write 
\begin{align}
	\mathbb E [g_j^2 ] &\equiv 
	\left\langle\left( \frac{\{\hat H,\hat  L_j\}}2 + \lambda_j \hat L_j\right)^2\right\rangle_{\hat \rho(\mb\theta)} \\&=
																																																										\nonumber
	\left\langle\left( \frac{\{\hat H,\hat  L_j\}}2 \right)^2+
	\lambda_j\frac{\{\hat L_j,\{\hat H,\hat  L_j\}\}}2 +
	\lambda_j^2\hat L_j^2\right\rangle_{\hat \rho(\mb\theta)} 
																																																										\\&=
																																																										\nonumber
	\left\langle\left( \frac{\{\hat H,\hat  L_j\}}2 \right)^2+
	\lambda_j\frac{\{\hat L_j,\{\hat H,\hat  L_j\}\}}2 
\right\rangle_{\hat \rho(\mb\theta)} +\lambda_j^2 {\rm QFI}_j~,
\end{align}
where $\{\hat A,\hat B\}=\hat A\hat B + \hat B \hat A$.
Since QFI is always positive, the optimal value of the ``baseline'' $\lambda_j$
is the vertex of the above parabola, namely
\begin{equation}
	\lambda^{\rm opt}_j = -\frac{\left\langle\{\hat L_j,\{\hat H,\hat  L_j\}\} \right\rangle_{\hat \rho(\mb\theta)}}{4 {\rm QFI}_j}~.
\end{equation}
We note that the bound \eqref{Fbound1} continues to hold even when the optimal
baseline is used, as by definition $\mathbb E [g_j^2 ]$ with 
the optimal baseline is smaller than $\mathbb E [g_j^2 ]$ for the non-optimal $\lambda_j=0$.

\section{Bound on $G_{\rm noisy}$}\label{a:bound2}
We first  focus on the estimator based on the log-derivative trick. We write the 
cost function as 
$C = \sum_y E_y  p(y|\mb\theta)$, where 
$p(y|\mb\theta) = \bra y \hat \rho(\mb\theta) \ket y $,  
$\hat H=\sum_y E_y \hat \Pi_y$ is the possibly unknown eigendecomposition of $H$ and
$\hat \Pi_y = \ket y\!\bra y$. 
Then 
\begin{equation}
	\nabla_j C = \mathbb{E}_{y\sim p(y|\mb\theta)}[ E_y \nabla_j \log p(y|\mb\theta)]~.
\end{equation}
From the above, we find that
$g_j = E_y \nabla_j \log p(y|\mb\theta)$ is an unbiased estimator of $\nabla_j C$. 
We recall the definition of the constants  $G_{\rm noisy}$ and $G_{\infty}$ 
such that
\begin{align}
	\mathbb E \left[\sum_j  g_j^2 \right] &\le G_{\rm noisy}^2~, &
	\max_j \mathbb E \left[g_j^2 \right] &\le G_\infty^2~.
	\label{upperG}
\end{align}
To get those constants 
 we need to find upper bounds for $\mathbb E \left[g_j^2 \right] $.
By explicit calculation, following a similar derivation of Ref.~\cite{paris2009quantum} we find
\begin{align}
	\mathbb E [g_j^2 ]  &= \sum_y  E_y^2 p(y|\mb\theta)  [\nabla_j \log p(y|\mb\theta) ]^2
									 \\ &= \sum_y  E_y^2 \frac{[\nabla_j p(y|\mb\theta) ]^2}{p(y|\mb\theta)}
									 \\ &\stackrel{(a)}{=} \sum_y  E_y^2 \frac{ [\Tr \hat \Pi_y(\hat \rho \hat L_j+\hat L_j\hat \rho)/2  ]^2}{\Tr[\hat \Pi_y \hat \rho]}
									 \\ &= \sum_y  E_y^2 \frac{ [\Re\Tr (\hat \Pi_y\hat \rho \hat L_j) ]^2}{\Tr[\hat \Pi_y\hat  \rho]}
									 \\ &\leq \sum_y  E_y^2 \frac{ \left|\Tr (\hat \Pi_y\hat \rho \hat L_j) \right|^2}{\Tr[\hat \Pi_y\hat  \rho]}
									 \\ &= \sum_y  E_y^2 
									 \left|\Tr \left(\frac{ \sqrt{\hat \Pi_y}\sqrt{ \hat \rho} }{\sqrt{\Tr[\hat \Pi_y\hat  \rho]}}\sqrt{\hat \rho}\hat  L_j\sqrt{\hat \Pi_y}\right) \right|^2
									 \\ &\stackrel{(b)}{\leq} \sum_y  E_y^2 
									 \Tr \left(\hat  \Pi_y \hat  L_j\hat \rho\hat  L_j\right) 
									 \\ &= 
									 \Tr \left(\hat  H^2 \hat  L_j\hat \rho\hat  L_j\right)~,
\end{align}
where 
in (a) we used the definition of the SLD \eqref{SLD}, and in (b) the Cauchy-Schwartz inequality. 
Using then the H\"older inequality and the fact that $\hat L_j\hat \rho \hat
L_j$ is a positive operator we find then
\begin{equation}
	\mathbb E [g_j^2 ]  \leq \|\hat H\|_\infty^2 \|\hat L_j\hat \rho\hat  L_j\|_1 \leq \|\hat H\|_\infty^2 {\rm QFI}_j ~,
	\label{Fbound1}
\end{equation}
where QFI$_j$ is the  Quantum Fisher Information \eqref{QFI}. The upper bounds \eqref{upperG} then follows with
\begin{align}
	G &= \|\hat H\|_\infty \sqrt{P\left(\max_j {\rm QFI}_j\right) }~, \\
	G_\infty &= \|\hat H\|_\infty \sqrt{\max_j {\rm QFI}_j}~.
\end{align}

A similar bound is obtained with another unbiased estimator of the gradient.
Here we set $\lambda_j=0$, while the general case is studied 
in the next section. Using the SLD we note that 
\begin{align}
	\nabla_j C &= \Tr[\hat H(\hat \rho \hat L_j+\hat L_j\hat \rho)/2] = \frac12 \left\langle \hat H \hat L_j+\hat L_j \hat H\right\rangle_{\hat \rho(\mb\theta)}
					\\ &\equiv  \left\langle \Re(\hat H \hat L_j)\right\rangle_{\hat \rho(\mb\theta)}~,
\end{align}
where $ \langle \hat A\rangle_{\hat\rho} = \Tr[\hat \rho\hat A]$,
$\Re[\hat A] \eqdef (\hat A+\hat A^\dagger)/2$, 
so the gradient can be estimated by quantum measurements of the operator $\Re(\hat H\hat  L_j)$. An upper bound is 
then obtained as 
\begin{align}
	\mathbb E [g_j^2 ] &\equiv 
	\left\langle \Re(\hat H\hat  L_j)^2\right\rangle_{\hat \rho(\mb\theta)} \\&\leq 
		 \left\langle \Re(\hat H\hat  L_j)^2+\Im(\hat H\hat  L_j)^2\right\rangle_{\hat \rho(\mb\theta)}
\\ &= \frac12 \Tr[\hat \rho(\hat L_j\hat H^2\hat L_j + \hat H\hat L_j^2 \hat H)]
\\ &= \frac12 \Tr[\hat L_j\hat \rho\hat  L_j(\hat H^2 + \hat L_j^{-1}\hat H\hat L_j^2 \hat H\hat L_j^{-1})]~,
\end{align}
where we have assumed that $\hat L_j^{-1}$ exists. Using again the H\"older inequality 
we get
\begin{align}
	\mathbb E [g_j^2 ] &\leq \frac12 \|\hat L_j\hat \rho\hat  L_j\|_1  
	\left(\|\hat H\|_\infty^2 + 
	\|\hat L_j^{-1}\hat H\hat L_j\|_\infty^2\right)
									\\ &\leq
					\|\hat H\|_\infty^2 {\rm QFI}_j~,
\end{align}
which is equivalent to Eq.~\eqref{Fbound1}.

\section{Fluctuating parameters}\label{a:fluct}

We consider an experimentally motivated noise model where the parameters $\theta_j$ 
cannot be tuned exactly. The lack of exact accuracy is modeled by a Gaussian noise 
with variance $\sigma_j^2$. This corresponds to the following 
substitution 
\begin{equation}
	\theta_j \to \mathcal N(\theta_j,\sigma_j^2)~,
\end{equation}
namely that the parameters are normally distributed around a mean value $\theta_j$
with variance $\sigma_j^2$. In the limit $\sigma_j\to0$ we recover the deterministic 
unitary operation \eqref{ansatz}. 
For $\sigma_j\neq 0$ we show that the above noise can be expressed into the form 
of Eq.~\eqref{noisyansatz}. 
We first note that 
\begin{align}
	\mathcal E_j^{\theta_j}[\hat \rho] &= \int d\vartheta \, \frac{e^{-\frac{(\vartheta-\theta_j)^2}{2\sigma_j^2}}}{\sqrt{2\pi\sigma_j^2}} 
	\,e^{-i\vartheta \hat X_j} \hat \rho e^{i\vartheta \hat X_j}
															\\& =  \mathcal D_j \circ \mathcal U_j^{\theta_j}[\hat \rho] 
															 \equiv  \mathcal U_j^{\theta_j}\circ\mathcal D_j[\hat \rho] ~,
\end{align}
where $\mathcal U_j^{\theta_j}[\hat \rho] = e^{-i\theta_j \hat X_j} \hat \rho e^{i\theta_j \hat X_j}$ is the 
noiseless gate and 
\begin{equation}
	\mathcal D_j[\hat \rho] = \int d\vartheta \, \frac{e^{-\frac{\vartheta^2}{2\sigma_j^2}}}{\sqrt{2\pi\sigma_j^2}} 
	\,e^{-i\vartheta \hat X_j} \hat \rho e^{i\vartheta \hat X_j}~,
	\label{depdef}
\end{equation}
is independent on $\theta_j$. 
To simplify our discussion we assume that $\hat X_j^2=\openone$. Although a more general form 
can also be obtained in other cases, any tensor product of Pauli matrices 
satisfies the constraint $\hat X_j^2=\openone$, so we believe that this restriction covers the 
most common gates that can be implemented in current NISQ devices. 
From series expansion it is simple to show that
\begin{equation}
	e^{-i\vartheta \hat X_j}[\hat \rho] 	e^{i\vartheta \hat X_j} = \hat \rho + \sin^2(\vartheta)(\hat X_j\hat \rho \hat X_j-\hat \rho) - \frac{i}2 \sin(2\vartheta)
	[\hat X_j,\hat \rho]~.
\end{equation}
Performing the integration in \eqref{depdef}
we get a dephasing-like channel, but with more general operators $\hat X_j$
\begin{equation}
	\mathcal D_j [\hat \rho] = (1-\eta_j)\hat \rho + \eta_j \hat X_j\hat \rho \hat X_j~,
	\label{dephasing}
\end{equation}
where 
\begin{equation}
	\eta_j = \frac{1-e^{-2\sigma_j^2}}2~.
\end{equation}
For $\sigma_j\to0$ we see that $\eta_j\to 0$ and $\mathcal D_j$ reduces to the identity channel. 

We have studied the effect of Gaussian fluctuations in the parameters of a QAOA as a function of 
of the noise rate $\eta_j\equiv \eta$. We found that the two terms in the bound \eqref{noisybound} 
display the same behaviour as observed in Fig.~\ref{fig:bounds}.

\section{Noise model and parameters} \label{a:noise}
In our numerical simulations we have used a custom noise model built 
up applying a depolarizing channel and thermal relaxation errors 
after each gate. For each of them three parameters have to be set: 
a dimensionless \textit{gate error}, connected with the 
parameter of the depolarizing channel that follows every single and 
two qubit gate, and the relaxation and decoherence times.  
In order to have a realistic noise model, in our numerical experiments relaxation and decoherence times are chosen to be different for each qubit, and gate error varies not only among qubits but also with the specific gate it is associated to. 
Indeed, all these parameters are taken from the available current calibration data of the various IBM quantum processors: after obtaining them, we  have constructed the noise model using the class \textit{NoiseModel} available at the moment of writing in \textit{Aer} library of Qiskit
\cite{Qiskit}, the quantum computing software development framework from IBM.\\
As we are interested to analyse how the algorithm performance depends  on noise ``strength'', in our numerical experiments we have simulated an increasing noise strength by multiplying every gate error value  originally obtained querying IBM calibration data by a factor $f={2, 4, 10}$, at the same time decreasing by the same factor $f$ every relaxation and decoherence time value (we remark that we scale up the full collection of original parameters obtained from the calibration data, so that a realistic noise model is preserved).\\
An additional set of parameters appearing in the noise model are gate times, i.e. the time required to apply the desired gate, during which relaxation and decoherence phenomena take place. Even gate time depends on the specific type of gate and on the qubit they are applied to: in all our simulations we left them unchanged, using the original values provided by IBM calibration data sets.  \\ 
In Table~\ref{tab:label} we report as a reference the average values for the IBMQ-16-Melbourne processor at the time of writing and running our simulations. 
The reader may refer to the Qiskit documentation \cite{Qiskit} for the explanation of the noise model and how it affects the different elementary gates.  \\
 
\begin{table}[htpb]
	\centering
	\caption{Parameters of the noise model}
	\label{tab:label}
	\begin{tabular}{ccc} 
			\hline
			\hline
			Gate & $\quad$& gate error \\ 
			\hline 
			U1 & $\quad$& $0$ (virtual gate) \\
			U2 &$\quad$&  $ 1 \cdot 10^{-3}$ \\
			U3 & $\quad$& $ 3 \cdot 10^{-3}$ \\
			CNOT &$\quad$&  $4 \cdot 10^{-2}$\\
			\hline 
			Relaxation time &$ \quad$ & Decoherence time\\ 
			$55\, \mu s$ &$ \quad$ & $68\, \mu s$ \\
			\hline 
			Single qubit gate time & $\quad $ & Two-qubit gate time\\
						$0.08\, \mu s$ & $\quad$ & $0.7\, \mu s$ \\
						\hline
		\end{tabular}
\end{table}

\section{Analytic gradient evaluation}
\label{a:gradient}
We use a classical gradient descent algorithm to minimize the cost function
\eqref{cost}. The information about the gradient of the cost function can be directly extracted by measuring the corresponding quantum observables: this procedure is often referred to as {\it analytically
evaluated gradient}, in the sense that we can analytically define a quantum circuit to estimate the gradient value for fixed parameters. Here we focus on the 
Hadamard test \cite{harrow2019low}.

To find the analytic gradient, 
we first denote with $\hat U_{m:n}$, with $m, n \in \{1,\dots, P\}, \, m\leq n$, the unitary
operator that applies the gates from  the $m$-th to the
$n$-th steps in the variational ansatz \eqref{ansatz}:
\begin{equation}
\label{Ujl}
\hat
U_{m:n}= e^{-i \theta_m \hat{X}_m}... e^{-i \theta_{n} \hat{X}_{n}}\, .
\end{equation} 
By deriving the cost function with respect to the $k$-th parameter, $\theta_k$,
$k \in \{1, \dots, P\} $, we obtain: 
\begin{equation}
\begin{split}
\frac{\partial C(\bm{\theta})}{\partial \theta_k} &= \bra{\psi_0}
\partial_{\theta_k} \hat U^{\dagger}_{1:P} \hat{H} \hat U_{1:P} \ket{\psi_0} +
\bra{\psi_0} \hat U^{\dagger}_{1:P} \hat{H} \partial_{\theta_k} \hat U_{1:P}
\ket{\psi_0}\\
&= -2 \Im \bra{\psi_0} \hat U^{\dagger}_{1:k} \hat{X}_k \hat U^{\dagger}_{k+1:P} \hat{H} \hat U_{1:P} \ket{\psi_0} \, ,
\end{split}
\end{equation}
where the last equality is obtained by using the definition \eqref{Ujl}. Using
the Pauli decompositions for both $\hat{X}_k$ and $\hat{H}$
\begin{equation}
\begin{cases}
\hat{X}_k = \sum_\mu \beta_\mu^{(k)} \hat Q_\mu^{(k)} \\
\hat{H}= \sum_\nu \alpha_\nu \hat P_\nu \, , 
\end{cases}
\end{equation}
where $\hat Q_\mu^{(k)}$ and $\hat P_\nu$ are Pauli operators, 
we can write the above derivative as 
\begin{equation}
\label{der1}
\frac{\partial C(\bm{\theta})}{\partial \theta_k}= - 2 \sum_\mu \sum_\nu
\beta^{(k)}_\mu \alpha_\nu \Im \bra{\psi_0} \hat U^{\dagger}_{1:k} Q_\mu^{(k)}
\hat U^{\dagger}_{k+1:P} \hat P_\nu \hat \hat U_{1:P} \ket{\psi_0} \, .
\end{equation}
Every term in the sum \eqref{der1} can be evaluated with a generalized Hadamard
test, that requires an additional ancilla qubit.
 The Hadamard test is performed with the following steps:
\begin{enumerate}
	\item initialize the ancilla qubit in the state $\ket{+}_A$ and the {\it principal register} 
		in the state $\ket{\psi_0}$;
	\item apply $\hat U_{1:k}$ to the principal register;
	\item apply $\hat Q_\mu^{(k)}$ to the principal register, controlled by the ancilla;
	\item apply $\hat U_{k+1:P}$ to the principal register;
	\item apply $\hat P_\nu$ to the principal register, controlled by the ancilla;
	\item apply a $\pi/2$ rotation around the $x$-axis to the ancilla and measure
		the latter on the computational basis.
\end{enumerate}
The probability of getting the outcome $0$ after the above steps is
proportional to $\Im \bra{\psi_0} \hat U^{\dagger}_{1:k} \hat Q_\mu^{(k)}
\hat U^{\dagger}_{k+1:P} \hat P_\nu \hat U_{1:P} \ket{\psi_0}$. Repeating the Hadamard test for
all $l$ and all $j$, and performing the sum expressed in \eqref{der1} we can
obtain an estimation of the analytic expression \eqref{der1} of the $k$-th derivative, so repeating all these 
steps for $k \in \{1...P\}$ we can evaluate the gradient.

\end{document}